\documentclass[aps,prl,twocolumn,superscriptaddress]{revtex4-1}

\usepackage{graphicx}
\usepackage{amsmath,amsfonts,amssymb}
\usepackage[colorlinks=true,linkcolor=blue,citecolor=blue]{hyperref}

\begin{document}

\title{Precursor configurations and post-rupture evolution of Ag-CO-Ag single-molecule junctions}

\author{Zolt{\'a}n~Balogh}
\affiliation{Department of Physics, Budapest University of Technology and Economics and MTA-BME Condensed Matter Research Group, \\ 1111 Budapest, Budafoki ut 8., Hungary}

\author{D{\'a}vid Visontai}
\affiliation{Physics Department, Lancaster University, LA1 4YB, Lancaster, United Kingdom}

\author{P{\'e}ter Makk}
\affiliation{Department of Physics, Budapest University of Technology and Economics and MTA-BME Condensed Matter Research Group, \\ 1111 Budapest, Budafoki ut 8., Hungary}

\author{Katalin Gillemot}
\affiliation{Physics Department, Lancaster University, LA1 4YB, Lancaster, United Kingdom}

\author{L{\'a}szl{\'o} Oroszl{\'a}ny}
\affiliation{Department of Theoretical Physics, Institute of Physics, Budapest University of Technology and Economics, 1111 Budapest, Budafoki ut. 8., Hungary}

\author{L{\'a}szl{\'o} P{\'o}sa}
\affiliation{Department of Physics, Budapest University of Technology and Economics and MTA-BME Condensed Matter Research Group, \\ 1111 Budapest, Budafoki ut 8., Hungary}

\author{Colin Lambert}
\affiliation{Physics Department, Lancaster University, LA1 4YB, Lancaster, United Kingdom}

\author{Andr{\'a}s Halbritter}
\affiliation{Department of Physics, Budapest University of Technology and Economics and MTA-BME Condensed Matter Research Group, \\ 1111 Budapest, Budafoki ut 8., Hungary}



\begin{abstract}
Experimental correlation analysis and first-principles theory are used to probe the structure and evolution  of Ag-CO-Ag single-molecule junctions, both \emph{before the formation, and after the rupture} of the junctions. Two dimensional correlation histograms and conditional histograms demonstrate that prior to the single-molecule bridge configuration the CO molecule is already bound parallel to the Ag single-atom contact. This molecular \emph{precursor configuration} is accompanied by the opening of additional conductance channels compared to the single-channel transport in pure Ag monoatomic junctions. To investigate the post-rupture evolution of the junction we introduce a cross-correlation analysis between the opening and the subsequent closing conductance traces. This analysis implies that the molecule is bound rigidly to the apex of one electrode, and so the same single-molecule configuration is re-established as the junction is closed. The experimental results are confirmed by ab initio simulations of the evolution of contact geometries, transmission eigenvalues and scattering wavefunctions.
\end{abstract}

\maketitle

\section{Introduction}

Break junction measurements play an essential role within the field of molecular electronics: after the rupture of a metallic nanowire an atomic-sized nanogap is created, which is a perfect tool for contacting single molecules and to studying their transport properties. \cite{agrait03,molelectr, kiguchi_review, Aradhya2013} Furthermore, using the break junction method a huge statistical ensemble of experiments is performed: the metallic nanowire is opened and then closed again several times, and from the thousands of conductance versus electrode separation traces a conductance histogram is created. The repeated formation of single molecule junctions is reflected by the growth of new peaks in the histogram in addition to the peaks of clean single- or a few atom configurations. \cite{xu03,smit02, lowTH2_1,Kiguchi1, PtCO, stat_anal, Bruot2012, Boer} It was shown that after the characterization of single-molecule junctions by conductance histograms, several more advanced data analysis methods could be applied to gain further information, including plateaus' length analysis,\cite{yanson98, Oren2, PtCO, lowTO2} two dimensional conductance-displacement histograms, \cite{RT3, Switch1, RT1, Mishchenko1, doi:10.1021/ja4015293, Zhou, ralph, riel} two-dimensional cross-correlation histograms, \cite{korrel2,korrel1, Venkata1,Hamill} or even the advanced investigation of  histogram-peak line-shapes. \cite{hersam1}

In this paper we address the additional  question of how one can investigate the evolution of the binding molecule \emph{prior to} the formation of a single-molecule junction, and \emph{after} junction rupture. To this end we investigate the interaction of carbon-monoxide molecules with silver atomic-sized junctions using two dimensional correlation histograms and conditional histograms, and we introduce a new type of correlation histogram based on the cross-correlation analysis of opening and subsequent closing traces.

Based on the correlation analysis of opening traces we find that the formation of the single-molecule junction can be foreseen from the appearance of a molecular precursor configuration \cite{Tsutsui,Yokota,Huisman}, which has additional open conductance channels compared to clean Ag single-atom contacts. This precursor configuration is associated with a CO molecule bound parallel to a dimer Ag single-atom contact. After this precursor configuration, single-molecule bridges are formed, which are identified by the conductance histogram itself. A \emph{final configuration conductance histogram} is capable of resolving the splitting of the single-molecule peak to two subpeaks, which are interpreted as the perpendicular/parallel orientation of the CO molecule with respect to the contact axis. Finally, the cross-correlation analysis of opening and subsequent closing traces implies that after junction rupture, the CO molecule stays rigidly bound to the apex of one electrode instead of leaving the junction or flipping to the side of the apex. The experiments are supported by ab initio simulations, which confirm both the interpretation of the molecular precursor configuration and the post-rupture evolution, i.e. the rigid binding of the molecule to the apex of one electrode.

\section{Results and Discussion}

\begin{figure}[!htb]
\begin{center}
\includegraphics[width=\columnwidth]{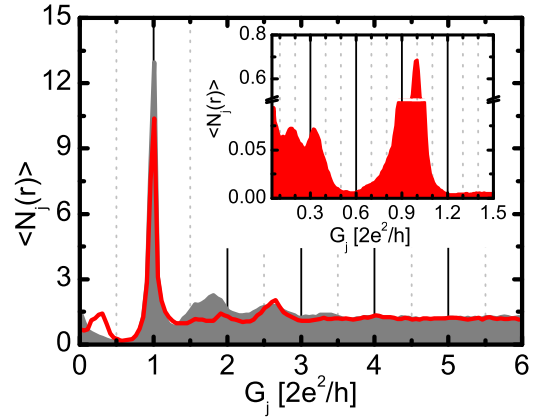}
\end{center}
\caption{\emph{Conductance histogram of pure Ag nanocontacts (grey histogram) and Ag-CO-Ag junctions (red curve) constructed from conductance traces along the repeated opening of the junction. The histograms are normalized to the number of traces. The new peak around 0.3$\,$G$_0$ is the sign of the Ag-CO-Ag molecular bridge formation. The inset shows the final configuration histogram of Ag-CO-Ag junctions (see text).}} \label{pull_hist.eps}
\end{figure}

First we analyze the conductance histogram of silver nanowires in a carbon-monoxide environment recorded using a low temperature mechanically controllable break junction device (see Methods for details). From the thousands of experimental conductance vs.\ electrode separation traces the histogram is calculated as $H_i=\left< N_i(r)\right>_r$, where $N_i(r)$ is the number of datapoints in the conductance bin $i$ on the r$^\mathrm{th}$ trace, and these numbers are  averaged over for the various traces. \cite{korrel2}
Figure \ref{pull_hist.eps} shows the conductance histogram of Ag junctions in a CO environment (red line) compared to the histogram of clean Ag junctions in high vacuum (grey histogram).
Compared to the pure Ag histogram \cite{venkata3,ludoph99,kiguchi2} the Ag-CO histogram shows a clear peak at $0.3\,$G$_0$, which is presumably related to Ag-CO-Ag single-molecule junctions (Fig.\ \ref{pull_hist.eps}). The peak at the conductance quantum unit ($1\,$G$_0$) seems similar for the pure and the molecular measurement, suggesting the formation of pure monoatomic Ag contacts in both cases.

The histogram peak of single-molecule junctions seems to exhibit a shoulder at $\approx 0.2\,$G$_0$. To further analyze this, we have constructed a \emph{final configuration histogram} (see inset in Fig.\ \ref{pull_hist.eps}), i.e. a conductance histogram for the last 10 datapoints of each conductance trace before the conductance jumps below a threshold of $\approx 0.01\,$G$_0$. According to our estimation this histogram is related to the final $\approx 1\,$\AA\ displacement before the contact rupture. The final configuration histogram shows a clear splitting of the single-molecule peak: two subpeaks are observed at $\approx 0.18\,$G$_0$ and $\approx 0.33\,$G$_0$, which we interpret as a perpendicular and a parallel CO molecule, like in Pt-CO-Pt single-molecule junctions.\cite{PtCO} In the traditional conductance histogram the finite slope of the single-molecule plateaus leads to the fusion of these two subpeaks.

The \emph{final configuration histogram} is also a proper tool to estimate the probability of molecular contact formation. According to the weight of the molecular and the single-atom region in the inset of Fig.\ \ref{pull_hist.eps}, one can state that approximately 20\% of all traces break from a single-molecule junction, whereas the rest of the junctions break from an Ag single-atom contact.

We note that the clear subpeaks in the histogram support the assumption that a single molecule is contacted: multiple molecules would lead to a broader  variety of contact geometries and a smearing of the histogram peak. Additionally, the small probability of molecule binding ($\approx$ 20\%) makes the formation of multiple molecule junctions unlikely.

\begin{figure}[!htb]
\begin{center}
\includegraphics[width=\columnwidth]{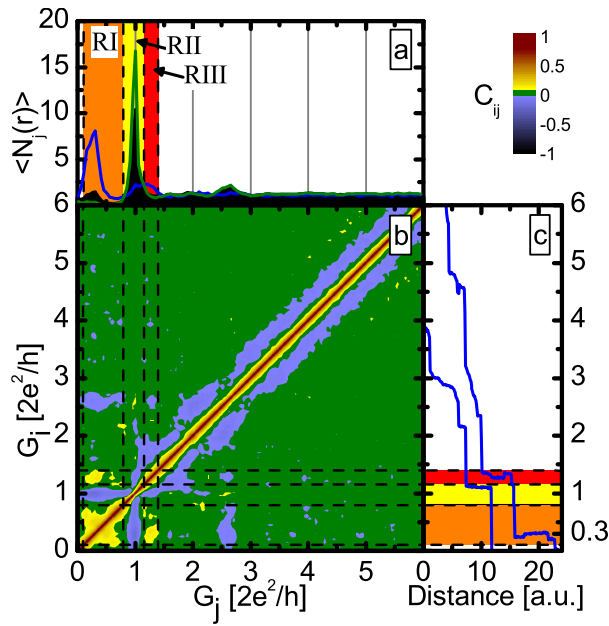}
\end{center}
\caption{\emph{a) Conductance histogram for all the traces (black histogram) and  conditional histogram for the molecular (RI) region (blue line) and the 1$\,G_{0}$ (RII) region (green line). b) 2D correlation histogram of Ag-CO junctions for opening traces. The negative correlation (cold colors) between the molecular region (RI) and the peak at 1$\,G_{0}$ (RII) is clearly visible. Between the interval of single-molecule configurations (RI) and the 1.15-1.4$\,G_{0}$ region (RIII) a positive correlation (warm colors) is observed.  c) Typical individual opening traces. The color rectangles and dashed lines show the three conductance region.}} \label{AgCO_hist-correl_pull.eps}
\end{figure}

As a next step we analyze the \emph{statistical relation} of the various contact configurations. To this end we apply the two dimensional correlation histogram (2DCH) technique introduced by a part of the authors in Refs.\citenum{korrel1,korrel2}. The 2DCH is defined as a two-dimensional correlation function:
\begin{equation}
C_{i,j}=\frac{\left\langle \left(N_{i}(r)-\left< N_{i}(r)\right> \right) * \left(N_{j}(r)-\left< N_{j}(r)\right> \right) \right\rangle }{\sqrt{\left\langle \left(N_{i}(r)-\left< N_{i}(r)\right> \right)^{2} \right\rangle \left\langle \left(N_{j}(r)-\left< N_{j}(r)\right> \right)^{2} \right\rangle }},
\label{corr.eq}
\end{equation}
where $N_{i/j}(r)$ is the number of datapoints on trace $r$ in the conductance bins $i$ or $j$, and all the averages are counted for the trace index $r$, similarly to the definition of conductance histograms. If on the conductance traces the number of datapoints at two different conductance intervals (around $G_i$ and $G_j$) are statistically independent from each other, then the correlation function is obviously zero. A distinct \emph{dependence}, however, introduces nonzero correlations.
A pair of contact configurations with conductances around $G_i$ and $G_j$ that usually \emph{appear together} (i.e. either both of them or none of them is observed on the same conductance trace) give rise to a positive correlation, $C_{i,j}>0$. On the other hand a negative correlation is observed if a larger than average weight of one configuration is accompanied by a smaller than average weight of the other, and vice versa. By definition $-1 \le C_{i,j}=C_{j,i} \le 1$. Further details about the correlation analysis are found in Ref.\citenum{korrel2}. With this method a correlation map can be drawn for arbitrary conductance pairs, $G_i$ and $G_j$.

Figure  \ref{AgCO_hist-correl_pull.eps}b shows the 2DCH for Ag-CO junctions. The two axes correspond to the two conductances, and the color shows $C_{i,j}$. Negative correlations are shown by cold colors (blue and black), positive by warm (yellow-red), and the green areas mark the regions, where the correlation is zero within the accuracy of the method. For further investigation we have defined three conductance regions, which are shown by different colors in panel a) and c), and are separated by black dashed lines on panel b). Region I (RI, orange) is related to the molecular configuration ($0.1\,$G$_{0}$-$0.8\,$G$_0$), the second interval (RII, yellow) corresponds to conductance of the single-atom Ag contacts ($0.8\,$G$_{0}$-$1.15\,$G$_0$) and the third (RIII, red) to the conductance region slightly above the single-atom peak ($1.15\,$G$_{0}$-$1.4\,$G$_0$).
Figure \ref{AgCO_hist-correl_pull.eps}b shows a clear anti-correlation between RI and RII and between RII and RIII. However, a positive correlation is observed between RI and RIII, which suggests that region III corresponds to a so-called \emph{precursor configuration} from which the single-molecule junction is likely to be formed. Optionally a pure Ag single-atom contact is formed, but then neither the precursor configuration, nor the single-molecule configuration is observed. These two optional trajectories are demonstrated by the two sample traces on panel c. Note, that the 2DCH does not resolve any clear distinction between the two molecular subpeaks demonstrated in the inset of Fig.\ \ref{pull_hist.eps}.

To obtain further justification we also study the correlations with conditional histogram technique, i.e. with histograms constructed only from those traces that have larger than average number of datapoints in a certain conductance interval.\cite{korrel1} In Fig.\ \ref{AgCO_hist-correl_pull.eps}a the blue curve represents the conditional histogram for RI, the green curve is the conditional histogram for RII, whereas the black histogram is the traditional histogram for the entire dataset. By definition all the histograms are normalized to the number of included traces.
Again, a clear anti-correlation is seen for the two selections: for the green histogram (selection for RII) the molecular peak disappears, whereas for the blue one (selected for RI) the peak at $1\,$G$_0$ disappears, and instead a new, wider peak appears around $1.3\,$G$_0$. This positive correlation was already seen in the 2DCH and corresponds to the precursor configuration of the molecular junction indicating that the CO molecule first binds
to the side of the Ag-Ag junction and by opening an additional transport channel increases the conductance from $1\,$G$_0$
to $1.3\,$G$_0$.

It is well known from conductance fluctuation measurements that pure Ag monoatomic junctions have a single, highly-transmitting conductance channel, and the further channels give negligible contribution to the conductance.\cite{ludoph99} This single-channel transport is related to conduction by solely $s$ electrons.\cite{cuevas98a} As a consequence the histogram of pure Ag shows a sharp upper edge of the single-atom peak at $1\,$G$_0$. With this knowledge it is clear that the precursor configuration -- which has a conductance significantly above $1\,$G$_0$ -- is definitely not a pure monoatomic Ag junction. Conceivably it is rather a monoatomic Ag junction with a CO molecule bound to the side. In the latter case the CO molecule may open up parallel channels for the transport, and so the conductance is no longer limited to $1\,$G$_0$.

It is to be emphasized that the molecular precursor configuration is hidden in the traditional conductance histogram, as it is suppressed by the sharp peak due to pure Ag monoatomic configurations. Similarly, the traditional conductance histogram does not reveal the fate of the molecule \emph{after} the rupture of the junction, which is analyzed in the next section.

\begin{figure}[!htb]
\begin{center}
\includegraphics[width=\columnwidth]{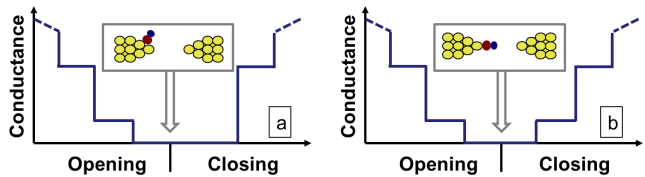}
\end{center}
\caption{\emph{Illustrative opening/closing conductance traces. a) The opening trace first shows a single-atom configuration (or a molecular precursor configuration), after which the plateau of the single-molecule junction is observed. After the rupture the molecule may flip to the side of the junction. In this case the closing trace does not show the single-molecule bridge, it immediately jumps to the single-atom/precursor configuration. b) If the molecule stays rigidly bound to the apex of one electrode after the rupture, then the closing trace  exhibits the same single-molecule plateau as the opening trace.}} \label{open_close}
\end{figure}

After rupture of the junction the molecule may either stay rigidly bound to the apex of one electrode tip, or alternatively it may flip to the side of the electrode, or even diffuse further away. These two situations can be distinguished based on the closing trace, i.e. by recording the conductance trace when the junction is closed after the complete rupture. If the molecule stays rigidly bound to the junction, it is likely that the closing trace will show the same molecular configuration as the opening trace. On the other hand, if the molecule leaves the junction, the closing trace will not show any molecular configuration (see illustration in Fig.\ref{open_close}).

To investigate this, we introduce a new type of correlation histogram based on the cross-correlation of the opening and the subsequent closing traces:
\begin{equation}
C'_{i,j}=\frac{\left \langle \left(N_{i}(r)-\left< N_{i}(r)\right> \right) * \left(N'_{j}(r)-\left< N'_{j}(r)\right> \right) \right \rangle }{\sqrt{\left\langle \left(N_{i}(r)-\left< N_{i}(r)\right> \right)^{2} \right\rangle \left\langle \left(N'_{j}(r)-\left< N'_{j}(r)\right> \right)^{2} \right\rangle }}.
\end{equation}
Here, $r$ denotes an entire opening--closing trace pair, $N_{i}(r)$ is the number of datapoints in bin $i$ within the opening part of the trace, whereas $N'_j(r)$ is the number of datapoints in bin $j$ within the closing part. In contrast to Eq.\ \ref{corr.eq} the diagonal of the opening--closing cross-correlation function is not unity, $C'_{i,i}\ne 1$, as $N_i(r)$ and $N'_i(r)$ are not related to the same part of a conductance trace. Therefore, a high positive value of $C'$ around the diagonal indicates that the closing part of the traces follows similar junction configurations as the opening part. Specifically for the Ag-CO-Ag single-molecule junctions a positive correlation is expected between RI of the opening part and the same conductance interval of the closing part (denoted by RI') if the molecule stays rigidly bound to the tip apex, whereas close to zero correlation is expected between RI and RI' if the molecule flips to the side after the contact rupture.

\begin{figure}[!htb]
\begin{center}
\includegraphics[width=\columnwidth]{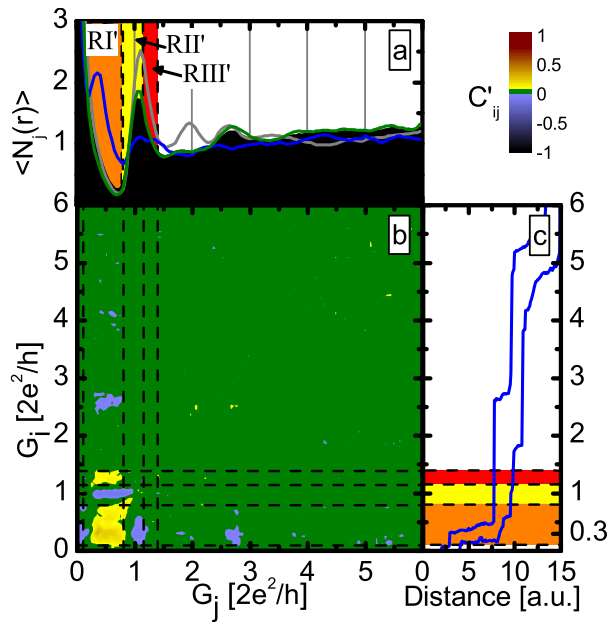}
\end{center}
\caption{\emph{a) Conditional histograms for closing traces with selection based on the prior opening process. Blue histogram is made for the traces, where a molecular junction was formed, for the green histogram the selection process was done to the 1$\,G_0$ region. The grey histogram is constructed for a clean Ag dataset, which was measured before the dosing of molecule. b) 2D correlation histogram of Ag-CO-Ag junctions addressing correlations between opening and closing traces. The vertical axis is associated with the opening, the horizontal axis with the closing process. c) Representative closing traces. }} \label{AgCO_hist-correl_push.eps}
\end{figure}

Fig.~\ref{AgCO_hist-correl_push.eps}b shows the correlation matrix obtained with the definition above. The vertical axis belongs to the opening, whereas the horizontal to the subsequent closing process. Going along a horizontal line gives the relation between a given configuration of the opening process and the various configurations along the subsequent closing process. The dashed lines mark the same conductance regions, as the ones in Fig~\ref{AgCO_hist-correl_pull.eps}. It is clear that the molecular configuration in the closing part (RI') positively correlates with the molecular configuration in the opening part (RI), whereas the single-atom configuration in the closing part (RII') is negatively correlated with the molecular configuration in the opening part (RI). The correlation diagram indicates, that if the molecular junction was formed during the opening process, then it is likely that it will be formed during the closing process as well, whereas if no sign of the presence of the molecules was seen during the opening, then this will likely be the case for the closing process too.

This is even better demonstrated by Fig~\ref{AgCO_hist-correl_push.eps}a, where the conditional histograms are plotted for the closing traces. Here the conditional histograms were selected according to the opening process: the blue line shows the histogram for the closing part of those traces, in which the molecular configuration (RI) has higher than average weight during the opening part; whereas the green curve shows the closing histogram for the traces with higher than average weight of the single-atom configuration (RII) along the opening part. As a reference the closing histogram for all Ag-CO traces (black histogram) and the closing histogram for a pure Ag dataset (grey) are also shown. The clear peak at $0.3\,$G$_0$ in the blue histogram again demonstrates, that whenever a single molecule Ag-CO-Ag junction is formed along the junction opening, it usually stays rigidly bound to one apex after the rupture, and so the molecular junction is re-established during the closing process. Note, that this feature is hidden in the closing histogram for the entire dataset (black).

\begin{figure}[!htb]
\begin{center}
\includegraphics[width=\columnwidth]{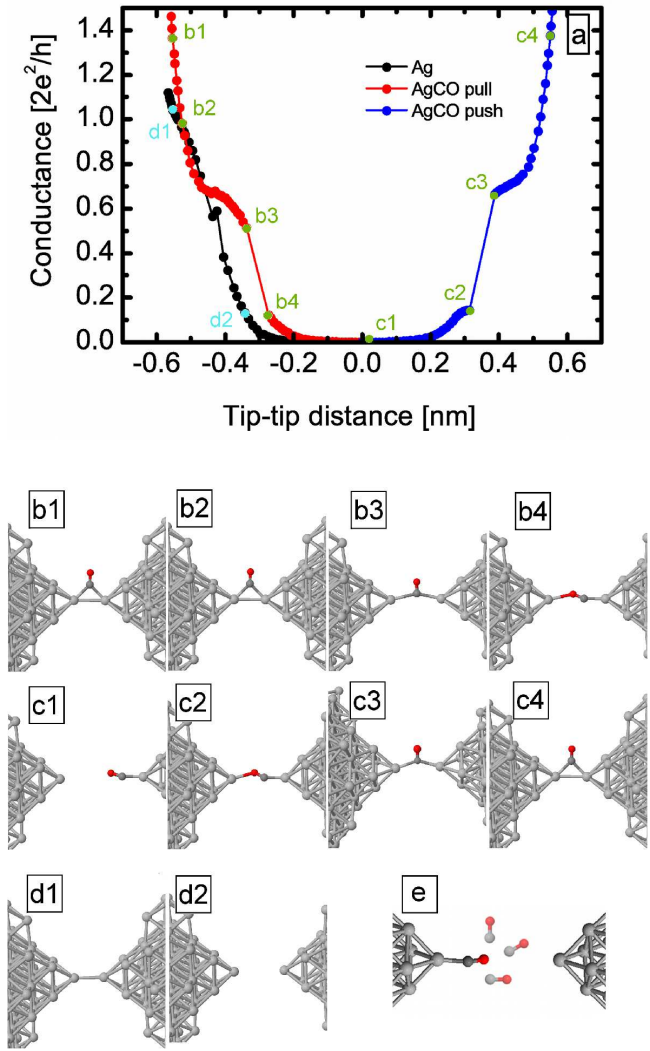}
\end{center}
\caption{\emph{Simulated opening (red) and closing (blue) trace of an Ag-CO-Ag junction and opening trace of a clean Ag junction (black) with the relevant geometries shown below the figure (b1-d2). e) CO molecule bound to one electrode in different starting configurations, from which it relaxes to the junction axis.}} \label{fig4}
\end{figure}

To verify the above interpretation of the experimental results, we performed ab initio simulations both of  junction evolution and  electronic transport properties (see Methods for details). To obtain starting configurations for initiating simulations of junction evolution, first we inserted the CO molecule close to an Ag dimer junction in various positions and orientations, and allowed the systems to relax.  Afterwards we have simulated the opening and closing traces of clean dimer-like Ag single atom contacts (Fig.\ \ref{fig4}d1-d2), and Ag-CO molecular junctions (Fig.\ \ref{fig4}b1-c4), by increasing or decreasing the electrode separation in steps of $0.01\,$nm and after each step relaxing the junction. The junction geometries illustrated in Fig.\ \ref{fig4}b1-c4 are examples of the stable configurations obtained from this procedure.

The theoretical opening and closing traces for the Ag-CO junction are shown in Figure~\ref{fig4}a as red and blue curves respectively, together with the geometries for selected points on the traces (see green points and the corresponding geometries under the figure). An extended plateau appears around $0.5-0.7\,$G$_0$, which corresponds to a configuration in which the CO molecule sits between the electrodes in a perpendicular configuration (b3), with the carbon atom in the binding axis. After further elongation the conductance shows a step like decrease when the molecule rotates into a parallel configuration as shown in Fig.~\ref{fig4}b4. The parallel configuration has an initial conductance of $\approx 0.12\,$G$_0$, which decreases to zero during further elongation.

The calculated conductance of the perpendicular/parallel configuration overestimates/underestimates the corresponding experimental peak positions in the inset of Fig.\ \ref{pull_hist.eps}. As the simulation is constrained to an idealized contact geometry, it is reasonable that the results of the simulations somewhat deviate from the experimental findings.\cite{colin1} It is possible that experimentally the CO molecule is not bound to an ideal Ag dimer, but to a less regular Ag structure.

After breaking of the junction, the molecule stays attached to the electrode and does not flip to the side. The closing trace, which is shown with blue on the right in Fig.~\ref{fig4}a shows similar behaviour to the opening traces; first the parallel configuration is formed and the CO molecule rotates from the parallel to the perpendicular configuration during further closing (see in Fig. \ref{fig4}c1-c3). Both the parallel and the perpendicular configurations have similar, but a bit higher conductance than in the opening process, as also seen in the experiment. This results from the smaller (unstretched) atom-atom distance which is formed during the closing process. To check the relevance of the rigid binding of the molecule to one apex, we rotated the molecule to various angles with respect to the contact axis, and afterwards we relaxed the system. For all the cases demonstrated in Fig.\ \ref{fig4}e, the molecule has relaxed towards the contact axis, which justifies that the molecule preferably stays in the contact axis.

\begin{figure}[!htb]
\begin{center}
\includegraphics[width=\columnwidth]{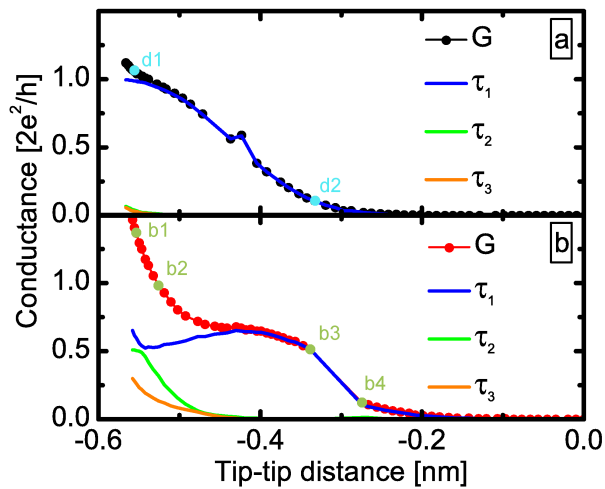}
\end{center}
\caption{\emph{Evolution of the transmission eigenvalues for the simulated opening traces of Fig.\ \ref{fig4}. a) For clean Ag junction the transport is dominated by a single channel. b) In the Ag-CO single-molecule junction two additional conductance channels open due to the binding of the CO molecule parallel to the Ag-Ag junction.}} \label{fig5}
\end{figure}

We also investigated the precursor effect with the help of the simulations: the black curve in Fig.~\ref{fig4}a shows the opening trace of the clean Ag dimer, which is now compared to the molecular trace. At small tip-tip distances, the conductance of the Ag-CO junction is higher than in the clean case.
The difference between an Ag single-atom contact and the molecular precursor configuration is even better demonstrated by the transmission eigenvalues of the simulated junctions (Fig.~\ref{fig5}). In accordance with experiment\cite{ludoph99} the clean Ag dimer has a single dominant conductance channel, and the remaining of the channels give a negligible contribution. However, the CO molecule bound to the side of the Ag--Ag dimer indeed opens up new channels: the molecular precursor configuration (see Fig.~\ref{fig5}b1) has three non-vanishing transmission eigenvalues. All these demonstrate that our previous interpretation of the precursor configuration as a CO molecule bound to the side of an Ag dimer is reasonable.

\begin{figure}[!htb]
\begin{center}
\includegraphics[width=0.8\columnwidth]{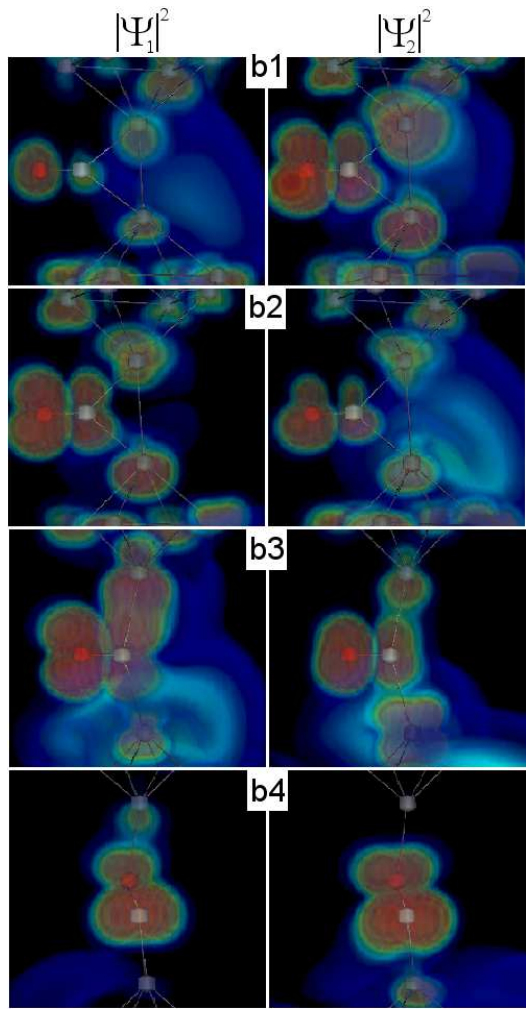}
\end{center}
\caption{\emph{Illustration of the the scattering wavefunctions for the first and second conductance channels related to the contact geometries b1-b4 in Fig.\ \ref{fig4}.}} \label{fig6}
\end{figure}

For further illustration we demonstrate the scattering wavefunctions of the first two conductance channels in Fig.\ \ref{fig6}. Usually the nature of these two channels are not well distinguishable, however for the b2 configuration it is clear that the first channel is related to the transport through the CO molecule bound to the side, whereas the second channel is rather related to the direct transport between the Ag tip atoms.

As the junction is further opened, the direct transport between the Ag tip atoms vanishes, and the current is dominated by the transport through the perpendicular/parallel CO molecule (Fig.\ \ref{fig6}b3/b4). In this case only a single conductance channel dominates the transport (Fig.\ \ref{fig5}b).

\section{Conclusions}

In conclusion, we have investigated the electronic and geometrical properties of the Ag-CO-Ag molecular junctions. As usual for these measurements, the formation of the molecular junction was demonstrated with conductance histogram measurements. However, by adopting a robust statistical analysis, which utilises correlation and conditional histograms, we have shown that precursor configurations and post-rupture evolution of the junction can be studied. These reveal that the CO molecule binds to the junction already before the formation of the Ag-CO-Ag single-molecule bridge configuration. This precursor configuration is related to a CO molecule bound parallel to the  Ag-Ag junction, which is accompanied by the opening of additional transport channels through the molecule. Moreover, by correlating the opening and closing traces, we have found, that after breaking the contact, the molecule stays rigidly bound to the apex of one electrode, it does not diffuse away or flip to the side of the junction, and so the same molecular configuration is re-established as the junction is closed. The interpretation of the molecular precursor configuration, and the rigid binding of the molecule to one electrode after the rupture were confirmed by theoretical calculations.

Our findings demonstrate, that the correlation analysis supplemented with the cross-correlation histograms between opening and closing traces supplies valuable information about precursor molecular configurations and the post-rupture evolution of single-molecule junctions. As these methods do not require special experimental conditions, they can be applied in any break junction measurement where conductance histograms are recorded.

\section{Acknowledgement} We acknowledge the financial support from the Hungarian Scientific Research Fund, OTKA K105735 and K108676 research grant. We are grateful to J. Ferrer, Sz. Csonka and J. Cserti for all the helpful discussions. We acknowledge financial support from the UK EPSRC, EP/K001507/1, EP/J014753/1, EP/H035818/1 and from the EU ITN MOLESCO 606728.

\section{Methods}
\textit{Experimental methods.}
The measurements were performed with a self-designed MCBJ setup at liquid helium temperature. High purity Ag wires ($0.125$mm diameter) were glued on the top of a phosphor-bronze bending beam by two drops of Stycast epoxy. Between the fixing points a notch was created reducing the diameter of the junctions to $10$-$30\,\mu$m. The fine tuning of the electrode separation was performed by a piezo actuator with a reduction of $\approx1:100$ due to the MCBJ geometry. The conductance versus electrode separation traces were measured by repeatedly opening and closing the junction with a symmetric triangle signal of $10$Hz on the piezo, and measuring the conductance through a current amplifier.

The precise dosing of CO molecules were performed with a home-made vacuum-system from a high purity container through a heated tube. The dosing method was done as follows: first we measured the clean Ag sample in cryogenic vacuum, after around $20'000-30' 000$ opening-closing cycles we heated the dosing tube to $90\,$K (above the CO boiling point) and a few "fake" dosing processes were done without opening the CO molecule container. With this procedure we could check the cleanness of our vacuum-system. After making sure, that the junction is not contaminated, and the dosing system is clean, we have done the real dosing procedure. The temperature of the tube was $90-100\,$K and we have observed signs of the molecule on the histogram after the dosage of $5\cdot 10^{-6}$mol of CO molecule. As the molecular peaks appear in the histogram the tube heating is turned off. At moderate bias voltage ($\approx 30-50\,$mV) the molecular peak is present for a long time without further dosing. Presumably the bias voltage slightly heats the junction, which aids the diffusion of the molecules on the junction surface. At elevated bias voltage ($\geq 200\,$mV) the molecules are desorbed, and the clean Ag histogram is reestablished.

\textit{Theoretical simulations}
Our ab-initio calculations were performed on the level of density functional theory as it is implemented in the code SIESTA \cite{elm1}. A double-zeta-polarized basis set was employed, with norm-conserving Troullier-Martins type non-relativistic pseudopotentials\cite{elm3} and an energy cutoff of $200$Ry to define the real space grid. The exchange and correlation energy was taken in to account  within the local density approximation (LDA) with the Perdew-Zunger parametrization \cite{elm2}.

$4$x$4$ $<001>$ Ag electrodes were constructed with pyramid shaped tips, then the system was relaxed for a large number of independent geometrical configurations, meaning different electrode separations and different starting positions for the CO molecule (e.g. different angle between the transport axis and the axis of the CO molecule, different distances between the junction and the CO), see Figure~\ref{fig4}e) for some examples. Except for the case when the CO was too far away from the junction, from all other initial positions the molecule always relaxed into the same local minimum corresponding to the specific electrode separation.

For the simulation in Fig.~\ref{fig4} we have chosen one of the over-squeezed initial geometries with lower tip-tip distance then the ideal Ag-Ag distance and simulated a full MCBJ opening and closing cycle, in which we consecutively relaxed the system and increased the electrode separation with $0.01$nm steps. To verify the first pulling curve, we then performed another pulling sequence, and found no deviation from the first cycle. The whole simulation was carried out for both when the CO molecule was present in the junction, and for the clean Ag-Ag junction as well.

Attaching 6 more layers of $<001>$ silver to each electrode, the Hamiltonian was calculated at each step of the above opening--closing cycle. The corresponding conductance traces, the channel decompositions and scattering wavefunctions were then obtained by the Gollum code \cite{elm4}.


\bibliographystyle{prsty}
\bibliography{AgCO}

\begin{thebibliography}{10}

\bibitem{agrait03}
N. Agra{\"\i}t, A.~L. Yeyati, and J.~M. van Ruitenbeek, Phys. Rep. {\bf 377},
  81  (2003).

\bibitem{molelectr}
J.~C. Cuevas and E. Scheer, {\em Molecular Electronics An introduction to
  Theory and Experiment} (World Scientific, Singapore, 2010).

\bibitem{kiguchi_review}
M. Kiguchi and S. Kaneko, Phys. Chem. Chem. Phys. {\bf 15},  2253  (2013).

\bibitem{Aradhya2013}
S.~V. Aradhya and L. Venkataraman, Nat Nano {\bf 8},  399  (2013).

\bibitem{xu03}
B. Xu and N.~J. Tao, Science {\bf 301},  1221  (2003).

\bibitem{smit02}
R.~H.~M. Smit {\it et~al.}, Nature {\bf 419},  906  (2002).

\bibitem{lowTH2_1}
S. Csonka, A. Halbritter, and G. Mih\'aly, Phys. Rev. B {\bf 73},  075405
  (2006).

\bibitem{Kiguchi1}
O. Tal {\it et~al.}, Phys. Rev. B {\bf 80},  085427  (2009).

\bibitem{PtCO}
P. Makk, Z. Balogh, S. Csonka, and A. Halbritter, Nanoscale {\bf 4},  4739
  (2012).

\bibitem{stat_anal}
M.~T. Gonz\'alez {\it et~al.}, Nano Letters {\bf 6},  2238  (2006).

\bibitem{Bruot2012}
C. Bruot, J. Hihath, and N. Tao, Nat Nano {\bf 7},  35  (2012).

\bibitem{Boer}
D. den Boer {\it et~al.}, The Journal of Physical Chemistry C {\bf 113},  15412
   (2009).

\bibitem{yanson98}
A.~I. Yanson {\it et~al.}, Nature {\bf 395},  783  (1998).

\bibitem{Oren2}
T. Yelin {\it et~al.}, Nano Letters {\bf 13},  1956  (2013).

\bibitem{lowTO2}
W.~H.~A. Thijssen, D. Marjenburgh, R.~H. Bremmer, and J.~M. van Ruitenbeek,
  Phys. Rev. Lett. {\bf 96},  026806  (2006).

\bibitem{RT3}
C.~A. Martin {\it et~al.}, Journal of the American Chemical Society {\bf 130},
  13198  (2008).

\bibitem{Switch1}
Y. Quek~Su {\it et~al.}, Nat Nano {\bf 4},  230  (2009).

\bibitem{RT1}
M.~T. Gonz\'{a}lez {\it et~al.}, J. Phys. Chem. C. {\bf 115},  17973  (2011).

\bibitem{Mishchenko1}
A. Mishchenko {\it et~al.}, Journal of the American Chemical Society {\bf 133},
   184  (2011).

\bibitem{doi:10.1021/ja4015293}
P. Moreno-Garc\'{i}a {\it et~al.}, Journal of the American Chemical Society
  {\bf 135},  12228  (2013).

\bibitem{Zhou}
X.-Y. Zhou {\it et~al.}, Nanoscale Research Letters {\bf 9},  77  (2014).

\bibitem{ralph}
E.~S. Tam {\it et~al.}, ACS Nano {\bf 5},  5115  (2011).

\bibitem{riel}
E. L\"{o}tscher {\it et~al.}, Small {\bf 9},  209  (2013).

\bibitem{korrel2}
P. Makk {\it et~al.}, ACS Nano {\bf 6},  3411  (2012).

\bibitem{korrel1}
A. Halbritter {\it et~al.}, Phys. Rev. Lett. {\bf 105},  266805  (2010).

\bibitem{Venkata1}
S.~V. Aradhya, M. Frei, A. Halbritter, and L. Venkataraman, ACS Nano {\bf 7},
  3706  (2013).

\bibitem{Hamill}
J.~M. Hamill, K. Wang, and B. Xu, Nanoscale {\bf 6},  5657  (2014).

\bibitem{hersam1}
M.~G. Reuter, M.~C. Hersam, T. Seideman, and M.~A. Ratner, Nano Letters {\bf
  12},  2243  (2012).

\bibitem{Tsutsui}
M. Tsutsui, M. Taniguchi, and T. Kawai, Nano Letters {\bf 9},  2433  (2009),
  pMID: 19507890.

\bibitem{Yokota}
K. Yokota, M. Taniguchi, M. Tsutsui, and T. Kawai, Journal of the American
  Chemical Society {\bf 132},  17364  (2010), pMID: 21086990.

\bibitem{Huisman}
E.~H. Huisman {\it et~al.}, Nano Letters {\bf 8},  3381  (2008), pMID:
  18771330.

\bibitem{venkata3}
T. Kim, H. V\'{a}zquez, M.~S. Hybertsen, and L. Venkataraman, Nano Letters {\bf
  13},  3358  (2013).

\bibitem{ludoph99}
B. Ludoph {\it et~al.}, Phys. Rev. Lett. {\bf 82},  1530  (1999).

\bibitem{kiguchi2}
S. Kaneko, T. Nakazumi, and M. Kiguchi, The Journal of Physical Chemistry
  Letters {\bf 1},  3520  (2010).

\bibitem{cuevas98a}
J.~C. Cuevas, A. {Levy Yeyati}, and A. Mart{\'\i}n-Rodero, Phys. Rev. Lett.
  {\bf 80},  1066  (1998).

\bibitem{colin1}
P. Makk {\it et~al.}, Phys. Rev. Lett. {\bf 107},  276801  (2011).

\bibitem{elm1}
J. Soler {\it et~al.}, J. Phys.: Condens. Matter {\bf 14},  2745  (2002).

\bibitem{elm3}
N. Troullier and J.~L. Martins, Phys. Rev. B {\bf 43},  1993  (1991).

\bibitem{elm2}
J.~P. Perdew and A. Zunger, Phys. Rev. B {\bf 23},  5048  (1981).

\bibitem{elm4}
J. Ferrer {\it et~al.}, Submitted  (2014).

\end{thebibliography}

\end{document}